\documentclass[aps,prl,floatfix,showpacs,reprint,superscriptaddress]{revtex4-1}
\usepackage{amssymb}
\usepackage{graphicx}
\usepackage{amsfonts}
\usepackage{amssymb}
\usepackage{natbib}
\usepackage{color}

\def\qq{{\rm q}}
\def\ss{{\rm s}}
\def\LL{{\cal L}}

\def\dd{{\rm d}}
\def\II{{\rm I}}

\def\aa{{\rm a}}
\def\bb{{\rm b}}

\begin{document}

\title{Shaping an Itinerant Quantum Field by Dissipation}

\author{Diego Porras}
\affiliation{
Departamento de F\'isica Te\'orica I,
Universidad Complutense, 
28040 Madrid, 
Spain}
\author{Juan Jos\'e Garc\'{\i}a-Ripoll}
\affiliation{Instituto de F\'{\i}sica Fundamental, IFF-CSIC, Serrano 113-B, 28006 Madrid, Spain}

\date{\today}

\begin{abstract}
We show that inducing sidebands in the emission of a single emitter into a one dimensional waveguide, together with a dissipative re-pumping process, a photon field is cooled down to a squeezed vacuum. Our method does not require to be in the strong coupling regime, works with a continuum of propagating field modes and it may lead to sources of tunable multimode squeezed light in circuit QED systems.
\end{abstract}

\maketitle

The quantum states of light are a crucial resource for precision measurements~\cite{xiao87prl} and quantum information processing~\cite{braunstein05rmp}. In the first case, single-mode squeezed states of the electromagnetic (EM) field make it possible to lower the uncertainty of measurements below quantum shot noise~\cite{xiao87prl}. In the second case, multimode squeezed states of continuous variables (the EM field) are the essential ingredient for a protocols to do quantum key distribution, teleportation, entanglement swapping, error correction and full fledged quantum computing~\cite{braunstein05rmp}. In particular, we remark the potential of transferring the entanglement from travelling multimode squeezed light to distant stationary qubits~\cite{kraus04prl}, to construct the backbone of quantum repeater protocols. In all these cases it is experimentally important to have robust and tunable sources of broadband squeezing, to
 overcome the inhomogeneous broadening of various experimental devices, such as qubits, detectors and passive and active elements.

This work shows how to create tunable continuous sources of single and multimode squeezed light by controlling single emitters coupled to propagating modes of the EM field. Our work builds on recent experiments that implement the main tools of cavity Quantum Electrodynamics (QED) using solid-state devices. These include superconducting qubits coupled to microwave transmission lines~\cite{wallraff04nat,*blais04pra,majer07nat}, as well as quantum dots coupled to microcavity photons~\cite{hennessy07nat}, or plasmons \cite{akimov07nat}. All those experiments combine the possibility of accurately controlling single quantum emitters by external fields, and coupling them to single-mode cavities or one-dimensional (1D) waveguides which support stationary or propagating modes of the EM field~\cite{bozhevolnyi06nat,houck07nat,astafiev10sci}. In particular we remark the maturity of the circuit-QED field, for which we will detail the actual physical implementation of our ideas.

The main results of this letter, presented in sequential order are: A multicolor driving of an artificial atom modifies its coupling to the EM field, inducing sidebands. Combining the sidebands with an auxiliary bath, a single qubit may cool a quantum field in a single mode cavity to a squeezed vacuum. If instead of a cavity, the driven qubit is placed in a waveguide, the high energy modes play the role of a dissipative bath and the result is tunable multimode squeezing of the propagating quantum field. Through the manuscript we will also discuss implementations, measurement schemes and further outlook.

\paragraph{Cooling to a photon squeezed vacuum.--} 
Let us introduce the general idea behind our main results. We will start with a single-mode photon field interacting with a qubit through $H_\II = g (a \sigma^+ + a^\dagger \sigma^-),$ where $a$ and $a^\dagger$ are the Fock operators for the field and $\sigma^{\pm}$ the qubit ladder operators. When the qubit has a very fast decay rate $\gamma_\qq \gg g$, it will cool the bosonic field to the bare photon vacuum, $|\Omega\rangle$, by a process in which the qubit continuously absorbs quanta of radiation and decays back to its ground state. Consider now that we engineer the qubit-field interaction to look like $H_\II = g(D \sigma^+ + D^\dagger \sigma^-)$, where $D =  u a + v a^\dagger$, $u^2 - v^2 = 1$. The photon field will now be cooled to a squeezed vacuum $|\Omega \rangle_{\rm s}$, determined by  the condition $D | \Omega \rangle_{\rm s} = 0$~\cite{walls.book,cirac92,*cirac93}. 
This process can be described by a Markovian evolution for the single-mode photon reduced density matrix, $\mu$, in terms of a Liouvillian superoperator,
\begin{equation}
\dot{\mu} = {\cal L}^{{[\rm s]}}_{\rm sq}(\mu) = \LL_{\{D, \Gamma^{\rm sq}\}}(\mu),
\label{cooling.single}
\end{equation}
with a parametrization of the Liouvillian in Linbladt form,
${\cal L}_{\{ O, \Gamma \} }(\mu) 
= \frac{\Gamma}{2} \left( O \mu O^\dagger \! - \! O^\dagger O \mu \right) 
 +  \mathrm{H.c.}$, where $\Gamma$ is the effective complex rate, and $O$ is the jump operator of the dissipative process.

In the second part of this work we will extend this idea to work with a continuum of bosonic modes, $a_\omega, a^\dagger_\omega.$ We will engineer a dissipative process that cools a 1D photon field to a multimode squeezed vacuum around two frequencies $\omega_{a,b}$, given by $D_\nu | \Omega \rangle_{\rm c} = \bar{D}_\nu | \Omega \rangle_{\rm c} = 0$, where the squeezing operators
\begin{equation}
D_\nu = u_\nu a_{\omega_\aa + \nu} + v_\nu a^\dagger_{\omega_{\bb} - \nu} , \
\bar{D}_\nu = {u}_\nu a_{\omega_\bb - \nu} + {v}_\nu a^\dagger_{\omega_\aa + \nu} ,
\label{multimode}
\end{equation}
satisfy $u_\nu^2-v_\nu^2 = 1$ and mutually commute $[D_\nu, \bar{D}_\nu] = 0.$ The multimode counterpart of Eq. (\ref{cooling.single}) is now
\begin{equation}
{\cal L}^{[{\rm c}]}_{\rm sq}(\mu) = \sum_{\nu} {\cal L}_{ \{ D_\nu, \Gamma^{\rm sq}_\nu \} } (\mu) + 
\sum_{\nu} {\cal L}_{ \{ \bar{D}_\nu, \bar{\Gamma}^{\rm sq}_\nu \} } (\mu).
\label{cooling.continuum}
\end{equation}

\paragraph{Photon sidebands by multicolor driving.--}
In order to implement the cooling dynamics (\ref{cooling.single}) and (\ref{cooling.continuum}) we need the ability to induce sidebands in the atom-field coupling, while controlling the qubit decay. We will now show how these sidebands may be achieved by simply driving the qubit energy levels, a method that is particularly suited for circuit QED and solid-state platforms.

Consider a qubit and a set of bosonic modes described by the free Hamiltonian
\begin{equation}
H_0 = \frac{\epsilon}{2} \sigma_z + \sum_\omega \omega a^\dagger_\omega a_\omega.
\end{equation}
The bare qubit-field interaction is given by
\begin{equation}
H_{\rm I} = \sum_\omega g_\omega (\sigma^+ + \sigma^-) (a_\omega + a_\omega^\dagger).  
\label{coupling}
\end{equation}
We will add a driving with one or more frequencies $\omega_{\dd,m}$ and relative amplitudes $\eta_{m}\;(m=1,2\ldots M),$
\begin{equation}
H_{\rm d}(t) = - \sum_m \eta_m \omega_{{\rm d},m} \cos(\omega_{{\rm d},m} t) \ \sigma_z.
\label{driving}
\end{equation}
In the interaction picture with respect to $H_0 + H_{\rm d}(t)$,
we get $H_{\rm I}(t) = \sigma^+ O(t) + {\rm H.c.}$, with
\begin{equation}
O(t)=\sum_\omega O_\omega(t) = \sum_\omega e^{i \epsilon t} G_\omega (t) (a_\omega e^{-i \omega t} 
+ a^\dagger_\omega e^{i \omega t}),
\label{jump}
\end{equation}
where $G_\omega(t)$ is an effective time-dependent coupling
\begin{equation}
G_\omega(t) = g_\omega
- g_\omega \sum_m \eta_m ( e^{i \omega_{\dd, m} t} - {\rm H.c.} )
+\mathcal{O}(\eta^2).
\label{G.omega}
\end{equation}
By setting $\omega_{\dd,m}  = \epsilon - \omega_m$ or $\omega_{\dd,m}= \epsilon +\omega_m$, we may resonantly select the survival of one or more sideband couplings, $\sigma^+a_{\omega_m}$ or $\sigma^+a_{\omega_m}$ in $H_{\mathrm{I}}(t)$, thus engineering the effective qubit-photon coupling.

\begin{figure}
\begin{center}
\includegraphics[width=\linewidth]{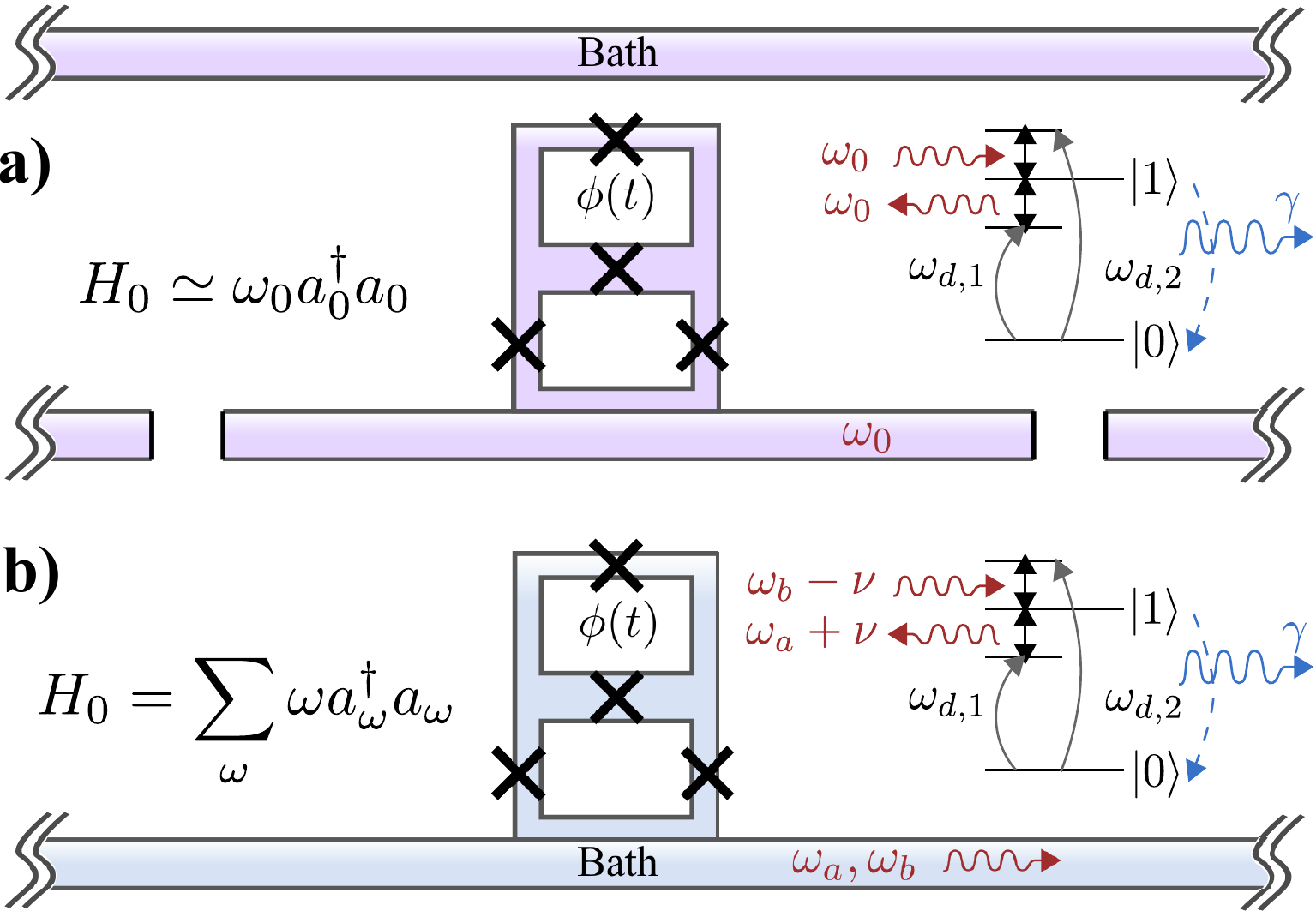}
\end{center}
\caption{
Scheme for shaping a quantum field by dissipation. (a) A flux qubit is coupled to a single mode microwave cavity. The qubit gap oscillates at frequencies $\omega_{d,1}$ and $\omega_{d,2}$ due to the flux driving of $\phi(t),$ and the qubit relaxes through the contact with the open line. (b) A similar qubit interacts with the 1D quantum field supported by a transmission line. High energy modes provide relaxation, while low energy-modes around $\omega_{\aa,\bb},$ evolve to a multi-mode squeezed vacuum.
}
\label{scheme}
\end{figure}

Let us introduce a possible experimental setup in which to test our ideas [Fig.~\ref{scheme}]. We will focus on a superconducting flux qubit where two junctions form a SQUID that allows the modulation of the qubit gap~\cite{paauw}. The external flux $\phi(t)$ applied on that loop will contain the multicolor driving that directly implements Eq.~(\ref{driving}). By the previous reasoning, the qubit coupling to the cavity [Fig.~\ref{scheme}a] or to the open line [Fig.~\ref{scheme}b] will be engineered to contain one or more sidebands. We will now discuss both cases and how they evolve into robust single- or multimode squeezing processes.

\paragraph{Single mode squeezing.--} We start with the particular case of a single mode cavity at frequency $\omega_0,$ $g_\omega = g \delta_{\omega , \omega_0}.$ By choosing $\omega_{\dd,1} = \epsilon - \omega_0$, $\omega_{\dd,2} = \epsilon + \omega_0$, we obtain the effective time-dependent jump operator (\ref{jump})
\begin{equation}
O(t) = \bar{g} D + \sum_{\lambda=1}^7 g_\lambda O_\lambda e^{- i E_\lambda t}+\mathcal{O}(\eta^2). \label{single.mode.dark}
\end{equation}
This contains the desired single-mode squeezing operator $D = u a_{\omega_0} + v a^\dagger_{\omega_0}$ with coupling strength $\bar{g} = g \left(\eta_1^2 - \eta_2^2 \right)^{1/2}$ and $u, v = \eta_{1,2}g/\bar{g}$, but in addition we find terms $O_\lambda=\{D^\dagger, a, a, a, a^\dagger, a^\dagger, a^\dagger\}$ that oscillate very rapidly with frequencies $E_\lambda = \{-2\epsilon, -\omega_{{\rm d},1}, -2 \omega_{{\rm d},1}, 2\omega_0, -\omega_{{\rm d},2}, -2 \omega_{{\rm d},2},-2\omega_0\}$ and amplitudes $g_\lambda=\{-\bar{g}, g, - \eta_1 g, \eta_2 g, g, - \eta_2 g, \eta_1 g\}.$

In order to get squeezed cooling we have to combine the sidebands with a fast decay of the qubit. In circuit-QED this can be engineered by approaching the qubit with an open transmission line that provides a relaxation channel~[Fig.~\ref{scheme}a]. Let us denote the qubit decay rate $\gamma_q.$ We will assume that the qubit is approximately in the ground state at all times $\rho = |0\rangle_{\rm q} \langle 0 | \otimes \mu,$ where $\mu$ is the photon reduced density matrix. Eliminating adiabatically $H_{\rm I}$ and keeping second order terms in the qubit-field coupling~\cite{breuer.book}, we get the effective time evolution,
\begin{equation}
d \mu / dt = \LL^{[{\rm s}]}_{\rm sq}(\mu) + \LL^{[{\rm s}]}_{\rm h/c}(\mu),
\label{eff.liouvillian}
\end{equation}
where the leading term is the squeezed cooling ${\LL}^{[\ss]}_{\rm sq}$ defined before~(\ref{cooling.single}). This equation is valid under the condition $\Gamma^{\rm sq} = 2 \bar{g}^2 / \gamma_{\rm q} \ll \gamma_\qq$, required for the adiabatic elimination of the qubit excited state. The residual terms lead to heating and cooling in the original basis, and appear with the usual photon losses in the cavity, of rate $\kappa$,
\begin{equation}
{\cal L}^{[{\rm s}]}_{\rm h/c} (\mu) 
= 
\sum_{\lambda=1}^7 {\cal L}_{ \{ O_\lambda,\Gamma^\lambda \} } (\mu)  
+ \LL_{\{ a, \kappa \}}(\mu).
\label{eq.heating}
\end{equation}
Since $\Gamma^{\lambda} = 2 g_\lambda^2 /(-i E_\lambda + \gamma)$, if we impose $\gamma_\qq \ll \omega_0, \omega_{{\rm d},\mu}, \epsilon$, which implies  $\Gamma^{\lambda}/\Gamma^{\mathrm{sq}} \approx (\gamma_\qq/E_\lambda)^2 \ll 1$, and if losses are small enough, $\kappa \ll \Gamma^{\rm sq}$,  then all corrections induced by ${\cal L}^{\mathrm{[s]}}_{\rm h/c}$ can be neglected.

\paragraph{Continuous spectrum.--}
We will now describe a process to engineer multimode squeezing of the EM field confined in an open 1D waveguide. The setup is very similar and consists of a ``bad'' qubit subject to multicolor driving and coupled to the a line of length $L$ ($L\to\infty$). However, two major technicalities arise. The first one is that the line now supports a continuum of modes, and in particular some of those modes may act as a ``bath'' for the qubit, providing the large $\gamma_q$ which we need. The second issue is that now we wish to perform multimode squeezing around two frequencies $\omega_a$ and $\omega_b$ of the spectrum [cf. Eq.~(\ref{multimode})]. For this we will need to implement a stroboscopic scheme that alternates two drivings of the qubit, generating both terms in Eq.~(\ref{cooling.continuum}).

We assume an Ohmic qubit-field coupling, $g_\omega = \sqrt{g_0 \ \omega}$, which describes superconducting qubits coupled to microwaves~\cite{blais04pra,wallraff04nat} and quantum dots in optical waveguides. The spectral density, $J(\epsilon) = \pi \sum_\omega g_\omega^2 \delta(\omega - \epsilon) = 2 \pi \alpha \omega$ determines a dimensionless coupling strength $\alpha,$ by which $g_0 = 2 \alpha \pi v /L=2\alpha\Delta\omega,$ where $v$ is the speed of light. In the weak coupling regime, $\alpha \ll 1$, the Born-Markov approximation is justified and a bare qubit with energy gap $\epsilon$ decays with a rate $J(\epsilon)$~\cite{leggett87rmp}. Since this rate grows with the energy, it makes sense to place our lossy qubit well above the frequencies that we want to squeeze, $\epsilon \gg \omega_{\aa,\bb}.$ This provides a separation of energy scales, where the high energy degrees of freedom act as a bath for the qubit and the low energy modes get squeezed.

More precisely, we choose a frequency cut--off $\omega_{\rm c}$, 
such that $\omega_\aa, \omega_\bb \ll \omega_{\rm c} \ll \epsilon$, and split the jump operator (\ref{jump}) 
$O(t) = O_{\rm low}(t) + O_{\rm high}(t)$ into 
$O_{\rm low}(t) = \sum_{\omega < \omega_{\rm c}} O_\omega(t)$, 
and $O_{\rm high}(t) = \sum_{\omega > \omega_{\rm c}} O_\omega(t)$. 
We define $\rho_{\rm low}$ as the reduced density matrix of the subsystem corresponding to low-energy frequency modes, plus the qubit. Tracing out the high energy modes and working in the interaction picture with respect to $H_0+H_{\mathrm{d}}(t)$, we get 
\begin{equation}
\frac{d \rho_{\rm low}}{d t} = {\cal L}_{\rm d}(\rho_{\rm low}) 
- i [\sigma^+ O_{\rm low}(t) + O^\dagger_{\rm low}(t) \sigma^- , \rho_{\rm low}],
\end{equation}
where
${\cal L}_{\rm d}
= {\cal L}_{ \{ \sigma^-, \gamma_{\rm q} \} }$, and 
$\gamma_{\rm q} = J(\epsilon) = 2 \pi \alpha \epsilon$ is decay rate of the qubit, to lowest order in the driving amplitude, $\eta_m$.

At this stage we are in a position similar to the qubit-cavity setup. The jump operator in the interaction picture contains the squeezing operator from Eq.~(\ref{multimode})
\begin{equation}
 O_{\rm low}(t) = \sum_{\nu } \bar{g}_{\nu} D_{\nu} e^{-i \nu t} + 
\sum_{\lambda, \omega} g_{\lambda, \omega} O_{\lambda,\omega} e^{-i E_{\lambda}(\omega) t} ,
\label{O.low}
\end{equation}
with $\bar{g}_{\nu}^2 = \eta_1^2 g_{\omega_\aa + \nu}^2 - \eta_2^2 g_{\omega_\bb - \nu}^2$ and $u_{\nu} = \eta_1 \ g_{\omega_\aa + \nu} / \bar{g}_{\nu}, v_{\nu} =
 \eta_2 \ g_{\omega_\bb - \nu} / \bar{g}_{\nu}$, but in addition we find fast rotating corrections $O_{1,\omega} = a_\omega$, $O_{2,\omega} = a^\dagger_{\omega}$, with $g_{1,\omega} = \eta_1 g_\omega$, $g_{2,\omega} = \eta_2 g_\omega$, and $E_1(\omega) = \omega + \omega_0$,  $E_2 (\omega) = - (\omega + \omega_0)$. Just like before, we now trace out the qubit, obtaining an evolution equation for the low-energy field modes density matrix,
$\mu = {\rm Tr}_{\rm q} \{ \rho_{\rm low} \}$, 
which to lowest order in the couplings $\dot{\mu} = {\cal L}^{[D_\nu]}_{\rm sq}(\mu) + {\cal L}^{[D_\nu]}_{\rm h/c}(\mu)$. The dominant term is obtained first in a time-dependent form,  $\propto \sum_{\nu,\nu'}\Gamma^{sq}_\nu e^{-i(\nu-\nu')t} \ldots$ with $\Gamma^{\rm sq}_{\nu} = \bar{g}_\nu^2 / (- i \nu + \gamma)$. Assuming a spectral resolution for the field modes larger than $\Gamma^{\rm sq}_\nu$, we can perform a rotating wave approximation, obtaining squeezing
\begin{equation}
\LL_{\rm sq}^{[D_\nu]} (\mu) \simeq
\sum_{\nu}
\Gamma^{\rm sq}_{\nu}
\left( D_{\nu} \mu D^{\dagger}_{\nu} 
- D^{\dagger}_{\nu} D_{\nu} \mu \right) + \mathrm{H.c.}.
\label{continuum.squeezing}
\end{equation}
together with corrections
\begin{equation}
\LL^{[D_\nu]}_{\rm h/c}(\mu) =
\sum_{\lambda, \omega} 
{\cal L}_{ \{ O_{\lambda,\omega}, \Gamma^\lambda_\omega \} } (\mu) 
\label{corrections}
\end{equation}
such as heating/cooling with $\Gamma_\omega^{\lambda} = g_{\lambda, \omega}^2/(-i E_\lambda(\omega) + \gamma_\qq)$, and potential photon losses, with rate $\kappa$. The effect of these corrections is negligible for modes around $\omega_\aa$, $\omega_\bb$ under conditions: (i) $\Gamma^\lambda_{\omega_\aa}, \Gamma^\lambda_{\omega_\bb} \ll \Gamma^{\rm sq}_0$, or equivalently, $\omega_\aa, \omega_\bb \ll \gamma_\qq$; and (ii), $\kappa \ll \Gamma^{\rm sq}_0$. 
Moreover, tracing out the qubit is justified if $\sum_\nu \Gamma^{\rm sq}_\nu \ll \gamma_\qq$.

\begin{figure}
\centering
\includegraphics[width = 3.3in]{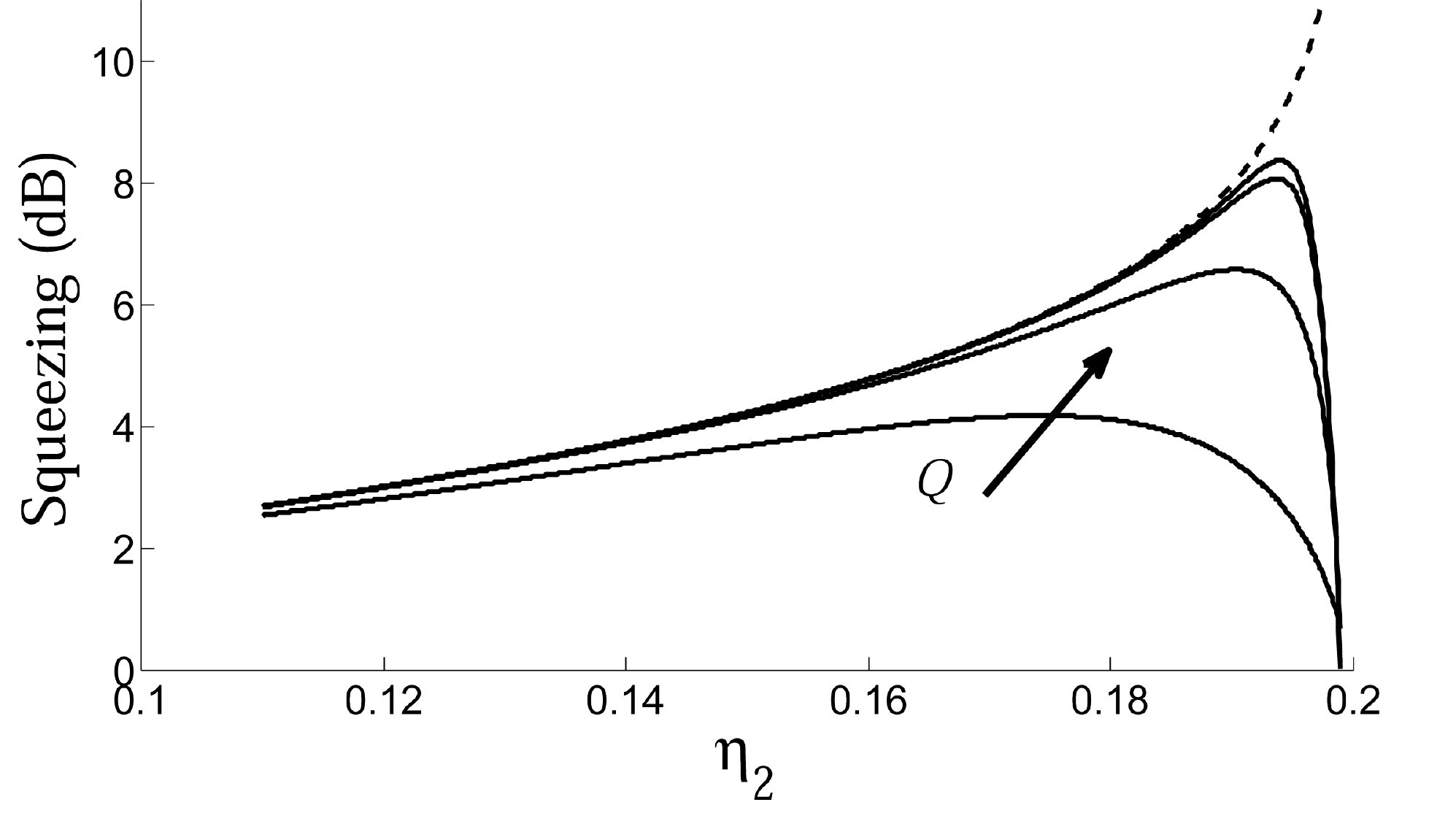}
\caption{Squeezing generated by dissipation in a single-mode microwave cavity: $\omega_0 = 3.5$, $\epsilon = 10$, $\gamma_\qq = 0.2$, $g=$ $1$  GHz, and $\eta_1 = 0.2$. 
We have used Eq.~(\ref{eff.liouvillian}) and the definition of squeezing in the main text. Continuous lines correspond to $Q = \omega_0/\kappa = 10^{5,6,7,8}$, and the dotted line is the ideal squeezed vacuum generated by $\LL_{\rm sq}^{[\rm s]}$.}
\label{squeezing}
\end{figure}

So far we have only implemented half of the squeezing process, the one that cools modes $D_\nu$ in Eq.~(\ref{multimode}). To cool also with the jump operators $\bar{D}_\nu$ we need another set of driving frequencies $\bar{\omega}_{{\rm d}, 1} = \epsilon - \omega_{\rm b}$, $\bar{\omega}_{{\rm d}, 2} = \epsilon + \omega_{\rm a}$, and driving amplitudes, $\bar{\eta}_1 = \eta_1 g_{\omega_{\rm a}} / g_{\omega_{\rm b}}$, $\bar{\eta}_2 = \eta_2 g_{\omega_{\rm b}} / g_{\omega_{\rm a}}$. In order to have \textit{both} cooling processes, in $D_\nu$ and $\bar{D}_\nu,$ we suggest using a stroboscopic cooling scheme, in which the system evolves during a time $t$ in $N$ cycles of duration $\Delta t  = t /(2 N)$, and driving parameters alternate between the ones associated to $D_\nu$ and those of $\bar{D}_\nu$. Choosing a small time interval $\Delta t \ll 1/\Gamma^{\rm sq}_{\nu, \nu}$, ensures the effective dynamics from Eq.~(\ref{cooling.continuum}) plus the same small corrections, which we already know how to neglect.

\begin{figure}
\centering
\includegraphics[width = 3.3in]{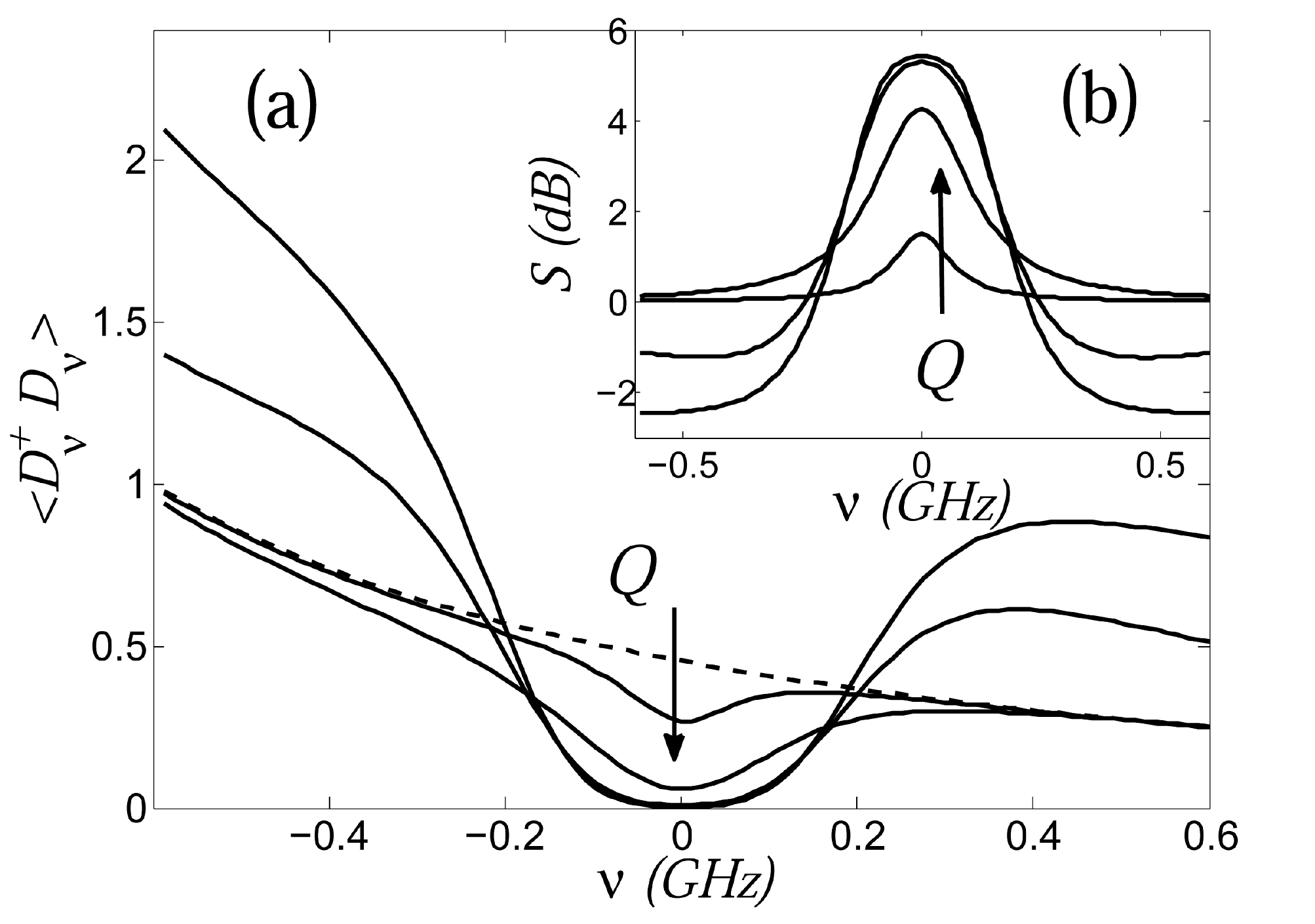}
\caption{Cooling of a one dimensional photon field to a squeezed vacuum. We have considered the stroboscopic method described in the text and a microwave cavity with $\epsilon$, $\omega_\aa$, $\omega_\bb$ $=$ 15, 3, 2.4 GHz, respectively. Qubit-field coupling strength corresponds to $\alpha =$ $6 \times 10^{-4}$, and quality factors are 
$Q = 10^{3, 4, 5, 6}$. (a) Occupation number in the basis defined by operators $D_\nu$, as a function of the frequency separation $\nu$ ($\nu = 0$, corresponds to squeezing between modes at $\omega_\aa$ and $\omega_\bb$). (b) Squeezing at each frequency is defined here as the suppression of quantum noise, $S = -10 \log_{10} \delta X_\nu$, for each two-mode quadrature 
$X_\nu = 
( a_{\omega_\aa + \nu} + a_{\omega_\bb - \nu} + \mathrm{H.c.} )/2$.
}
\label{continuous}
\end{figure}

\paragraph{Performance.--} To quantify the squeezing generated by the dissipative process, we use the steady state solution of Eqs.~(\ref{cooling.single}) and (\ref{cooling.continuum}), including the heating/cooling corrections due to fast rotating terms, Eqs.~(\ref{eff.liouvillian}) and (\ref{corrections}). For the single mode we define the quadrature $X = a + a^\dagger$, and relate squeezing to the suppression of quantum noise, $S = - \log_{10} \delta X $, where $\delta X^2 = \langle X^2 \rangle - \langle X \rangle^2$ is the variance of $X$ and $\delta X = 1$ for the EM vacuum. Fig.~\ref{squeezing} shows how squeezing improves with the cavity quality factor, and that ratios of 6 dB are attainable for realistic parameters. In the multimode case, shown in Fig.~\ref{continuous}, squeezing has to be defined with respect to a combined quadrature. 
In the continuum we estimate $\kappa = \Delta \omega / Q$, 
derived under the assumption that losses mainly happen at the boundaries of a long 1D waveguide, and using values of $Q$ describing small cavities in the single-mode limit. 
Remarkably, for weak coupling strengths, it is possible to cool a broad range of frequencies around $\omega_{a,b}$ to an entangled state with no excitations. The range of frequencies ($2.4-3$ GHz) is high enough to neglect thermal excitations, and the squeezing is large enough to be detected using a frequency-dependent variant of the cross-correlation methods devised in various groups~\cite{menzel10,eichler11}.

\paragraph{Conclusions and Outlook.--} 
We have shown that the photon quantum field of a cavity may be shaped by dissipation following a scheme that is ideally suited for circuit QED platforms. In the case of single mode cavities, our scheme requires an auxiliary bad cavity to induce dissipation, whereas in a 1D long waveguide, this is not required, since high energy photons play the role of a dissipative bath. Our proposal could allow experimentalist to control a continuous quantum field with the aid of a dissipative cooling process and generate tunable multimode squeezing. Other methods for generating sidebands may be applied to get the required qubit field couplings \cite{nori07prb}. 
Finally, our scheme may be easily generalized to other physical setups, like quantum dots coupled to photonic or plasmonic cavities. Here sidebands may be induced by using excited levels of charged dots, and by using polarized light to tune transitions between spin-states~\cite{imamoglu}. 

\bibliographystyle{apsrev}
\bibliography{cqed}

\end{document}